\documentclass[twocolumn, switch]{article}
\usepackage{preprint}
\usepackage[numbers,square]{natbib}

\usepackage{amsmath}
\usepackage{amssymb}
\usepackage{dcolumn}
\usepackage{graphicx}
\usepackage{bm}
\usepackage{braket}
\usepackage{multirow}

\usepackage[utf8]{inputenc}	
\usepackage[T1]{fontenc}
\usepackage{xcolor}
\usepackage{lineno}					
\usepackage{tikz} 					

\usepackage{newfloat}
\DeclareFloatingEnvironment[name={Supplementary Figure}]{suppfigure}
\usepackage{sidecap}
\sidecaptionvpos{figure}{c}

\usepackage{titlesec}
\titlespacing\section{0pt}{12pt plus 3pt minus 3pt}{1pt plus 1pt minus 1pt}
\titlespacing\subsection{0pt}{10pt plus 3pt minus 3pt}{1pt plus 1pt minus 1pt}
\titlespacing\subsubsection{0pt}{8pt plus 3pt minus 3pt}{1pt plus 1pt minus 1pt}

\definecolor{lime}{HTML}{A6CE39}

\title{Strategy of satellite QKD with passive high brightness entangled photon pair source}

\usepackage{authblk}

\author[1]{Jin-Woo Kim}

\author[1]{Suseong Lim}

\author[1]{Heonoh Kim}

\author[1*]{June-Koo Kevin Rhee}

\affil[1]{The School of Electrical Engineering, KAIST, Republic of Korea\protect\\
*rhee.jk@kaist.edu}

\begin{document}

\twocolumn[\begin{@twocolumnfalse}

\maketitle

\begin{abstract}
    A high-brightness entangled photon pair (HBEPP) source is essential for conducting entanglement-based quantum key distribution (QKD) between a satellite and a ground station. While an ultrabright source can overcome significant losses in satellite-based QKD (SQKD) and increase the sifted key rate, it also induces the multi-photon effect, raising the system's error rate. To accurately estimate system performance, we first present an analytical model for calculating the measurement probabilities of HBEPP distribution through an asymmetric loss channel. Based on this model, we propose the use of a passive-intensity HBEPP source for SQKD systems, assuming a polarization-independent channel and threshold detectors for measurement. We confirm that fixing the mean photon number at $\bar{\mu}=0.1$ achieves a performance of $99.7\%$ compared to the ideal one-way communication entanglement-based SQKD protocol, which is effectively optimizing the HBEPP source brightness in accordance with system losses.
\end{abstract}

\keywords{Entanglement, Multi-photon, BBM92, Satellite QKD}

\vspace{0.5cm}

\end{@twocolumnfalse}]


\section{Introduction}\label{sec1}
Quantum key distribution (QKD) is a technology that enables the secure distribution of a key between two users (Alice and Bob) by utilizing principles of quantum mechanics, ensuring safety from eavesdropping. Inspired by the concept of \textit{conjugate coding} proposed by S. Wiesner\cite{wiesner1983conjugate}, the introduction of the BB84 protocol by C. Bennett and G. Brassard opened new horizon for QKD using 
a single-photon source\cite{BENNETT20147}. However, practical photon sources do not emit single photons but instead generate multiple photons, making it difficult to assert the security of real systems based on the no-cloning theorem and the principles of quantum uncertainty. Fortunately, various research approaches in the field of QKD security have made it possible to ensure security even over long distances using actual equipments \cite{bennett1995generalized, shor2000simple, waks2002security, hwang2003quantum, gottesman2004security}.

The method of performing QKD by transmitting photons through optical fibers offers the advantage of easy system configuration; however, it must tolerate an attenuation channel of approximately 0.2 dB/km, making it unsuitable for long-distance transmission. Consequently, recognizing that transmitting photons through free space experiences relatively less attenuation, the design of satellite-based QKD (SQKD) was proposed early in the history of QKD, with experiments beginning shortly thereafter \cite{jacobs1996quantum,buttler1998free}. Although atmospheric effects on free-space systems for SQKD do not significantly impact transmission efficiency, diffraction associated with long-distance transmission still results in an attenuation of about 20 to 45 dB \cite{takenaka2017satellite, liao2017satellite, yin2017satellite, yin2020entanglement}. Even though QKD can be implemented over long distances for secure key distribution, the commercialization of such systems remains challenging if Alice and Bob are unable to generate a sufficient number of secure keys for communication. Consequently, it is imperative that QKD systems operating in high-attenuation environments require high-brightness sources. In the case of prepare-and-measure QKD protocols such as BB84, it has been proposed that utilizing decoy states enables secure QKD construction even over long distances with practical photon sources \cite{hwang2003quantum}. However, for QKD protocols such as BBM92, which rely on entanglement to ensure security, preparing a high-brightness entangled photon pair (HBEPP) source is more challenging than in BB84 systems.

The distribution of entanglement plays a crucial role not only in QKD systems but also in many quantum information technologies, including quantum networks. As a result, preparing HBEPP sources has become a key area of research. However, the challenge with HBEPP sources is that they exhibit multi-photon effects, which degrade the quality of the source due to the presence of unwanted polarization states. Since the initial exploration and development of HBEPP by P. Kwiat \cite{kwiat1995new}, theoretical analyses of the multi-photon effects in HBEPP sources have advanced \cite{simon2003theory, durkin2004resilience, caminati2006nonseparable, ma2007quantum, takesue2010effects, semenov2011fake, yoshizawa2012evaluation, takeoka2015full, brewster2021quantum}.Subsequently, experimental evidence has demonstrated that HBEPP sources can be effectively prepared for various applications in quantum information technologies \cite{caminati2006nonseparable, eisenberg2004quantum, kim2006phase, gerhardt2011experimentally, steinlechner2014efficient, cao2018bell}. However, previous theoretical analyses have either been applicable only to specific scenarios \cite{durkin2004resilience, ma2007quantum, takeoka2015full}, assumed conditions where losses are so significant that only single-photon level measurements are feasible \cite{caminati2006nonseparable, takesue2010effects}, or relied on numerical simulations \cite{yoshizawa2012evaluation}, without fully generalizing the issues. Recently, analyses of the distribution of HBEPP under general conditions have been conducted using the Fusimi-Kano method \cite{brewster2021quantum}. However, since this method is not yet widely known, cross-validation is needed.

Despite the limitations of existing analyses, theoretical tools for QKD analysis have been well established, and researchers who are now ready to develop high-brightness sources are now prepared to conduct long-distance QKD experiments outdoors. Initially, only BB84 experiments were conducted using weak coherent pulses\cite{hughes2002practical, kurtsiefer2002step}, but following the demonstration of the distribution of HBEPP over a distance of 600m in an outdoor\cite{aspelmeyer2003long}, several attempts have been made to perform BBM92 between two widely separated fixed locations\cite{peng2005experimental, marcikic2006free, ursin2007entanglement, liao2017long, ecker2021strategies, mishra2022bbm92}. Although these experiments did not address whether security could be guaranteed while operating BBM92 system under the influence of multi-photon effects, the performance of QKD remained consistent with the predictions presented by X. Ma \cite{ma2007quantum}. Subsequently, conditional security proofs were provided \cite{makarov2006effects, beaudry2008Squashing, tsurumaru2008security, kravtsov2023security}, which further accelerated the development of SQKD technology.

The technologies required to perform long-distance QKD in free space remain applicable for conducting the SQKD system, but they are not sufficient on their own. For a low earth orbit satellite moving along its orbit, tracking technology to maintain a stable link between the satellite and the ground station is also crucial. These technologies enable the conduct of mobility-based QKD experiments using hot air balloons \cite{wang2013direct}, airplanes \cite{nauerth2013air}, and pickup trucks \cite{bourgoin2015free}. Additionally, analyses of QBER through polarized laser beam signals between the satellite and the ground station have been conducted \cite{vallone2016interference, takenaka2017satellite}. A series of research activities and outcomes have sufficiently enabled the execution of SQKD \cite{liao2017satellite, yin2017satellite, yin2020entanglement}, marking a significant milestone in the history of the QKD field.

In this paper, we first analyze the distribution of entangled photon pairs to enable applications not only in QKD but also in various quantum information technologies, such as Bell tests and quantum teleportation. We present a general analysis model that considers the case where the measurement bases of Alice and Bob, who receive the entangled photon pairs, differ by an arbitrary angle and where their channels have asymmetric losses. Here, we make two assumptions. The first assumption is that the attenuation in the channel is independent of the direction of polarization, which is reasonable based on the characteristics of atmospheric conditions in free space. The second assumption is that a threshold detector is used for measurement, as placing a number-resolving detector on the satellite would not be cost-effective or spatially efficient for satellite QKD.

The newly presented analytical model starts from the Hamiltonian of  entangled photon pair generation \cite{simon2003theory, durkin2004resilience} and models the loss channel using a beam splitter (BS) to derive the distributed quantum states \cite{caminati2006nonseparable, takesue2010effects, yoshizawa2012evaluation}. We assume a general situation where the losses in Alice's and Bob's channels are asymmetric. The analytic solutions to the problem provide faster and accurate predictions for experimental models. In this paper, we calculate the probabilities of measurement outcomes that may occur in the system. From the results, we confirm that the CHSH value obtained matches the value calculated using the Fusimi-Kano method \cite{brewster2021quantum}. Additionally, we cross-validate that if the effects of multi-photon generation are not appropriately processed during post-processing, the CHSH value results in a fake violation that shows no-signaling outcomes \cite{semenov2011fake}.

The appropriate post-processing method for handling the multi-photon effects arising from HBEPP is referred to as the Squash model \cite{tsurumaru2008security, kravtsov2023security}. In this paper, we indicate that threshold detectors and the Squash model are adopted for the analysis of SQKD. Traditionally, QKD analysis optimizes the mean photon number of the source in relation to losses, and the system must be configured accordingly \cite{ma2007quantum}. However, due to the nature of SQKD, where the distance between the satellite and the ground station changes rapidly and losses fluctuate quickly, adjusting the mean photon number in real time based on measurements or predictions can be quite burdensome. Moreover, there is no guarantee that non-passive optical components will function properly when deploying the satellite; additionally, incorporating further systems within the limited resources of the satellite may not be efficient. Therefore, we analyzed changes in the mean photon number of the optimized entangled photon pair source within the loss range that low earth orbit satellites can accommodate. Consequently, we confirmed that the required optimized mean photon number does not change significantly within the range of losses encountered by the satellite, and the differences in the expected key rates are also minimal. Through this analysis, we propose fixing the mean photon number of the HBEPP utilized in SQKD.

\section{Generation and distribution of entangled photon pairs}\label{sec2}
First, we aim to generate polarization-based entangled photon pairs using spontaneous parametric down-conversion (SPDC) and a Sagnac interferometer, and to derive the quantum states by considering attenuation in the environmental channel from a quantum information perspective. When the output modes of the Sagnac interferometer are indexed as 1 and 2, the Hamiltonian for the generation of entangled photon pairs is given as \cite{simon2003theory, durkin2004resilience}, 
\begin{equation}\label{eq:Hamiltonian}
    \hat{H} = i\kappa\left(\hat{a}_{1H}^\dagger\hat{a}_{2V}^\dagger-
    \hat{a}_{1V}^\dagger\hat{a}_{2H}^\dagger\right) + h.c.
\end{equation}
Here, the subscripts of the creation operator $\hat{a}^{\dagger}_{ij}$ indicate whether it corresponds to the first mode (signal) or the second mode (idler), with $i\in\left\{1,2\right\}$, while $j\in\left\{H,V\right\}$ specifies whether the polarization of each mode is horizontal or vertical. The constant $\kappa$ is the nonlinearity constant that affects the rate of entangled photon pair generation, and $h.c.$ represents the Hermitian conjugate.

Considering the time evolution $U = e^{-i\hat{H}t/\hbar}$, the quantum state of the entangled photon pairs that accounts for the multi-photon effect is
\begin{equation}\label{eq:Psi}
    \ket{\Psi} = (1-g^2)\sum_{n=0}^{\infty}{\sqrt{n+1}g^n
    \ket{\psi_{-}^{n}}},
\end{equation}
where 
\begin{equation}\label{eq:psin-2}
     \ket{\psi_{-}^{n}} 
     = \frac{1}{n!\sqrt{n+1}}\left(
     \hat{a}_{1H}^\dagger\hat{a}_{2V}^\dagger-
    \hat{a}_{1V}^\dagger\hat{a}_{2H}^\dagger
     \right)^n \ket{0}.
\end{equation}
Here, $\gamma \equiv \kappa T_{int}$ denotes the gain of the nonlinear process in the generation of entangled photon pairs, where $T_{int}$ is the interaction time. Defining the nonlinear gain $g = \tanh \gamma$, the mean photon number within a single temporal mode is given by $\mu = \sinh^2{\gamma}$.

\begin{figure}
    \centering
    \includegraphics[width=\linewidth]{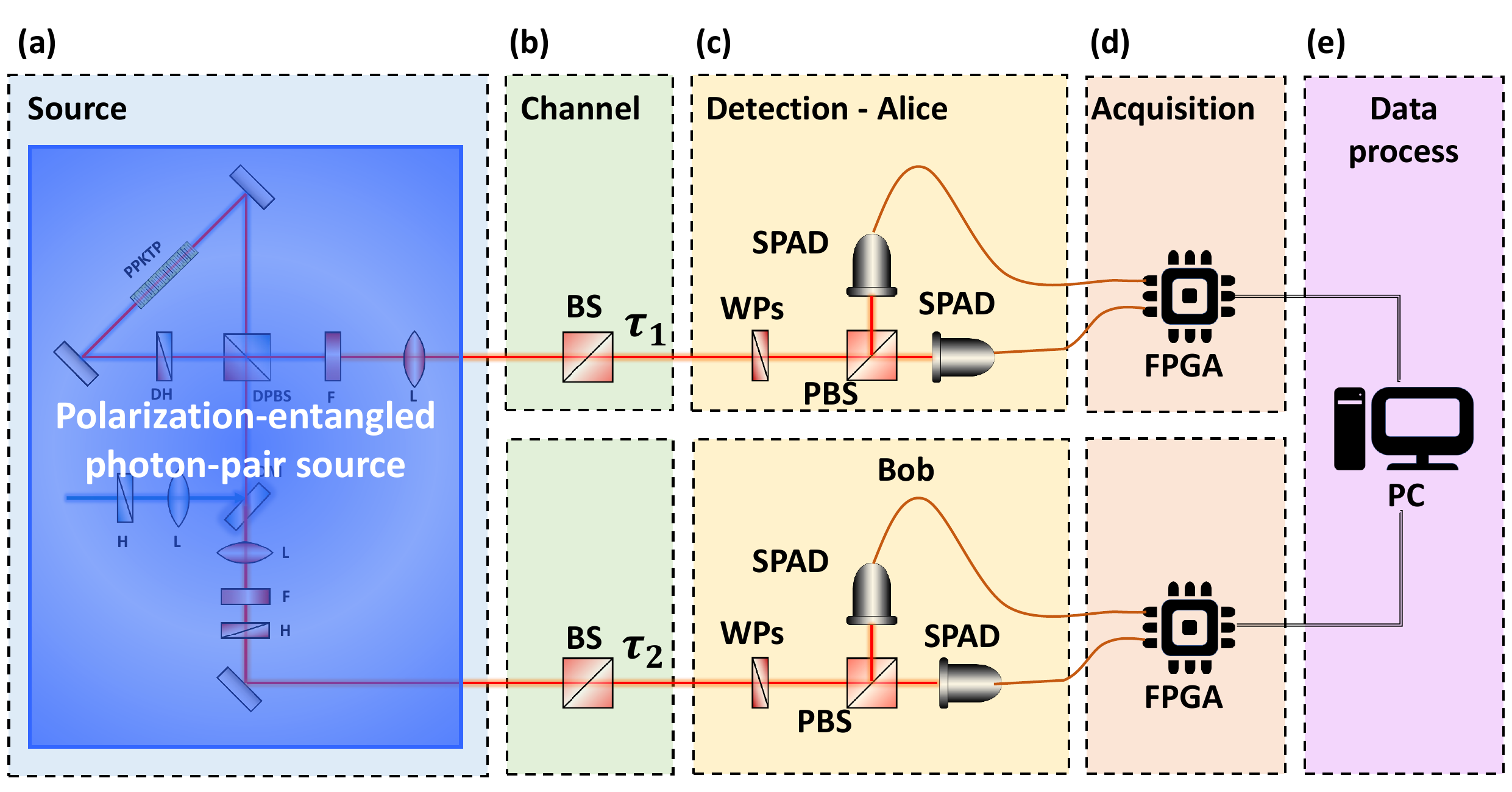}
    \caption{Overview of the modeling for long-distance entangled photon pair distribution. (a) The generation of the entangled photon pair source due to SPDC. The quantum state at this time is represented by equation (\ref{eq:Psi}). (b) Each transmittance $\tau_1$ and $\tau_2$ represents for channel of Alice and Bob, respectively. (c) Alice and Bob determine their measurement bases using wave plates(WPs) and polarization beam splitter(PBS). (d) Alice and Bob store the detection timing, detection basis, and detection results in high-speed field programmable gate array (FPGA). (e) Post-processing is performed to analyze the results of the distributed entangled photon pairs.}
    \label{fig:Scheme}
\end{figure}

To distribute the entanglement, the signal mode and idler mode of the entangled photon pair source are allocated to Alice and Bob, respectively, as shown in Fig. \ref{fig:Scheme}.
Since the entangled states distributed to Alice and Bob will be measured in bases defined by angles $\theta_1$ and $\theta_2$, respectively, it is useful to express the state in terms of these measurement bases. Accordingly, we define $\hat{a}^{\dagger}_{\theta^+_i} \equiv \cos\theta_i\hat{a}^{\dagger}_{iH}+\sin\theta_i\hat{a}^{\dagger}_{iV}$ and $\hat{a}^{\dagger}_{\theta^-_i} \equiv -\sin\theta_i\hat{a}^{\dagger}_{iH}+\cos\theta_i\hat{a}^{\dagger}_{iV}$. By substituting these expressions into Eq. (\ref{eq:psin-2}), we can reformulate $\ket{\psi_-^n}$ in terms of the relative measurement basis angle between Alice and Bob, which is given by $\theta \equiv \theta_1 - \theta_2$.
 
The attenuation that the quantum state experiences in the channel can be described using a BS model\cite{caminati2006nonseparable}. Given an input mode $\mathbf{k}_{in}$ and vacuum mode $\mathbf{k}_{vac}$, the BS model can be expressed as
\begin{equation}
    \begin{pmatrix}
        \hat{a}^{\dagger}_{out}\\
        \hat{a}^{\dagger}_{loss}
    \end{pmatrix}
    = 
    \begin{pmatrix}
        \sqrt{\tau_i} & i\sqrt{1-\tau_i}\\
        i\sqrt{1-\tau_i} & \sqrt{\tau_i}
    \end{pmatrix}
    \begin{pmatrix}
        \hat{a}^{\dagger}_{in}\\
        \hat{a}^{\dagger}_{vac}
    \end{pmatrix},
\end{equation}
where $\mathbf{k}_{out}$ and $\mathbf{k}_{loss}$ represent the mode of output state and the mode of the attenuated quantum state that is partially traced out, respectively. We assume that the cahnnel's attenuation acts independently of the polarization of the quantum state; therefore, we can represent the attenuated quantum state using the given transmittance $\tau_1$ and $\tau_2$ uniformly for all $\theta$ with respect to $\hat{a}^{\dagger}_{\theta_i^{\pm}}$. This allows us to obtain the final quantum state distributed to Alice and Bob, defined as $\rho_a = \text{Tr}_{loss}[\rho_{\ket{\Psi}\bra{\Psi}}]$\footnote{See Supplemental Material at URL-will-be-inserted-by-publisher for deriving the quantum state $\rho_a$.}. By utilizing number-resolving detectors, we can easily calculate the probabilities of each measurement result by obtaining the diagonal terms of $\rho_a$. However, in this paper, we analyze a more practical scenario by considering a threshold detector scheme, such as cost-effective Single Photon Avalanche Diodes (SPADs).

The commonly used Si-based SPAD exhibits a detection efficiency of about $70\%$. This detection efficiency can be included in the transmittance of the BS model. Therefore, it is possible to assume that each detector has a perfect detection efficiency. Now the measurement operator of a threshold detector for an arbitrary single mode $j$ is given as\cite{yoshizawa2012evaluation},
\begin{equation}\label{eq:detection}
    M_j \equiv d_x\ket{0}\bra{0}_j + \sum_{n=1}^{\infty}{\ket{n}\bra{n}_j}.
\end{equation}
Here, $d_x$ represents the rate of dark counts. From Eq. (\ref{eq:detection}), we can define the detection not only for single counts but also for n-fold coincidences. For example, the coincidence detection operator, where Alice detects at $\theta_1^+$ and Bob detects at $\theta_2^-$, is defined as $M=\sum_{n,m=1}^{\infty}{\ket{n,m}\bra{n,m}_{\theta_1^+,\theta_2^-}\ket{0}\bra{0}_{\theta_1^-,\theta_2^+}}$. In this case, to investigate the pure effects of multi-photon generation, we assume the case where dark counts are ignored, setting $d_x=0$. However, in actual experiments, dark counts can significantly degrade the performance of entanglement distribution and must be carefully considered.

We note that the above discussion is limited to a single temporal mode, requiring that the coincidence time window be set shorter than the coherence time\cite{yin2017satellite}. Even though the actual experiment is not conducted under single mode conditions, the results presented in this paper can be extended to multi-mode analysis. Therefore, for convenience, we assume that analysis is performed in a single temporal mode with high-speed time stamping for information storage. Alice and Bob can now store the results obtained from their respective detectors at each time interval. In our scheme, where Alice and Bob each use two detectors, there are a total of 16 possible outcomes.

\section{Results prediction for entanglement distribution}
The probabilities for each of the 16 possible outcomes, which can exist as a result of the measurements, can be obtained using the quantum state $\rho_a$ and the operators corresponding to each measurement outcome. The measurement probabilities are specified in Table. (\ref{tab:prob}).  Although previous works have attempted to determine the measurement probabilities for distributed quantum states \cite{caminati2006nonseparable, takesue2010effects, yoshizawa2012evaluation}, their approaches involve solving summations that include Fisher's noncentral hypergeometric distributions. Solving such problems is generally challenging; however, we have demonstrated that by applying Riemann’s theorem on removable singularities, it is possible to bypass these difficulties and compute the measurement probabilities more easily\footnote{For a detailed derivation of the probabilities for each of the 16 cases, see the Supplemental Material at URL-will-be-inserted-by-publisher.}.
\begin{table}[]
\caption{Alice and Bob each have two detectors for modes of $\theta_{1/2}^{\pm}$ and record the corresponding data on FPGA. If a detector clicks, it is recorded as 1; if it does not click, it is recorded as 0.}
\renewcommand{\arraystretch}{1.2} 
\resizebox{\columnwidth}{!}{
\begin{tabular}{c cc cc c c cc cc}
\hline\hline
\multirow{2}{*}{\textbf{Probability}} & \multicolumn{2}{c}{\textbf{Alice}}          & \multicolumn{2}{c}{\textbf{Bob}}            &  & \multirow{2}{*}
{\textbf{Probability}} & \multicolumn{2}{c}{\textbf{Alice}}          & \multicolumn{2}{c}{\textbf{Bob}}            \\ \cline{2-5} \cline{8-11} 
                                      & \multicolumn{1}{c}{\textbf{$\theta_1^+$}} & \textbf{$\theta_1^-$} & \multicolumn{1}{c}{\textbf{$\theta_2^+$}} & \textbf{$\theta_2^-$} &  &                                       & \multicolumn{1}{c}{\textbf{$\theta_1^+$}} & \textbf{$\theta_1^-$} & \multicolumn{1}{c}{\textbf{$\theta_2^+$}} & \textbf{$\theta_2^-$} \\ \cline{1-5} \cline{7-11} 
\textbf{$P_{0,0}$}                           & \multicolumn{1}{c}{0}          & 0          & \multicolumn{1}{c}{0}          & 0          &  & \textbf{$P_{\theta_1^-,\theta_2^-}$}                           & \multicolumn{1}{c}{0}          & 1          & \multicolumn{1}{c}{0}          & 1          \\[1.2ex] \cline{1-5} \cline{7-11} 
\textbf{$P_{\theta_1^+,0}$}                           & \multicolumn{1}{c}{1}          & 0          & \multicolumn{1}{c}{0}          & 0          &  & \textbf{$P_{\theta_1^+\theta_1^-,0}$}                           & \multicolumn{1}{c}{1}          & 1          & \multicolumn{1}{c}{0}          & 0          \\[1.2ex] \cline{1-5} \cline{7-11} 
\textbf{$P_{\theta_1^-,0}$}                           & \multicolumn{1}{c}{0}          & 1          & \multicolumn{1}{c}{0}          & 0          &  & \textbf{$P_{0,\theta_2^+\theta_2^-}$}                          & \multicolumn{1}{c}{0}          & 0          & \multicolumn{1}{c}{1}          & 1          \\[1.2ex] \cline{1-5} \cline{7-11} 
\textbf{$P_{\theta_2^+,0}$}                           & \multicolumn{1}{c}{0}          & 0          & \multicolumn{1}{c}{1}          & 0          &  & \textbf{$P_{\theta_1^+\theta_1^-,\theta_2^+}$}                          & \multicolumn{1}{c}{1}          & 1          & \multicolumn{1}{c}{1}          & 0          \\[1.2ex] \cline{1-5} \cline{7-11} 
\textbf{$P_{\theta_2^-,0}$}                           & \multicolumn{1}{c}{0}          & 0          & \multicolumn{1}{c}{0}          & 1          &  & \textbf{$P_{\theta_1^+\theta_1^-,\theta_2^-}$}                          & \multicolumn{1}{c}{1}          & 1          & \multicolumn{1}{c}{0}          & 1          \\[1.2ex] \cline{1-5} \cline{7-11} 
\textbf{$P_{\theta_1^+,\theta_2^+}$}                           & \multicolumn{1}{c}{1}          & 0          & \multicolumn{1}{c}{1}          & 0          &  & \textbf{$P_{\theta_1^+,\theta_2^+\theta_2^-}$}                          & \multicolumn{1}{c}{1}          & 0          & \multicolumn{1}{c}{1}          & 1          \\[1.2ex] \cline{1-5} \cline{7-11} 
\textbf{$P_{\theta_1^+,\theta_2^-}$}                           & \multicolumn{1}{c}{1}          & 0          & \multicolumn{1}{c}{0}          & 1          &  & \textbf{$P_{\theta_1^-,\theta_2^+\theta_2^-}$}                          & \multicolumn{1}{c}{0}          & 1          & \multicolumn{1}{c}{1}          & 1          \\[1.2ex] \cline{1-5} \cline{7-11} 
\textbf{$P_{\theta_1^-,\theta_2^+}$}                           & \multicolumn{1}{c}{0}          & 1          & \multicolumn{1}{c}{1}          & 0          &  & \textbf{$P_{\theta_1^+\theta_1^-,\theta_2^+\theta_2^-}$}                          & \multicolumn{1}{c}{1}          & 1          & \multicolumn{1}{c}{1}          & 1          \\[1.2ex]
\hline\hline
\end{tabular}
}\label{tab:prob}
\end{table}

The measurement probabilities for the distributed quantum state are expressed as a linear combination of $Q_j(\theta)$ for natural numbers $j\in[1,16]$. Here, $Q_j(\theta)$ is given by the following equation:
\begin{equation}\label{eq:Q-func}
Q_j=\frac{ABCD}{ABCD+G^2-G(AD+BC)\cos^2\theta-G(AC+BD)\sin^2\theta},
\end{equation}
where $G\equiv g^2(1-\tau_1)(1-\tau_2)$, and parameters are $A,B\in\left\{1,1-\tau_1\right\}$ and $C,D\in\left\{1,1-\tau_2\right\}$, satisfying the conditions that describe a total of 16 probability values\footnote{See Appendix A for the details}. Fig. \ref{fig:theta} shows how these measurement probabilities vary with the angle difference $\theta$ between Alice and Bob. 
\begin{figure}[t]
    \centering
    \includegraphics[width=\linewidth]{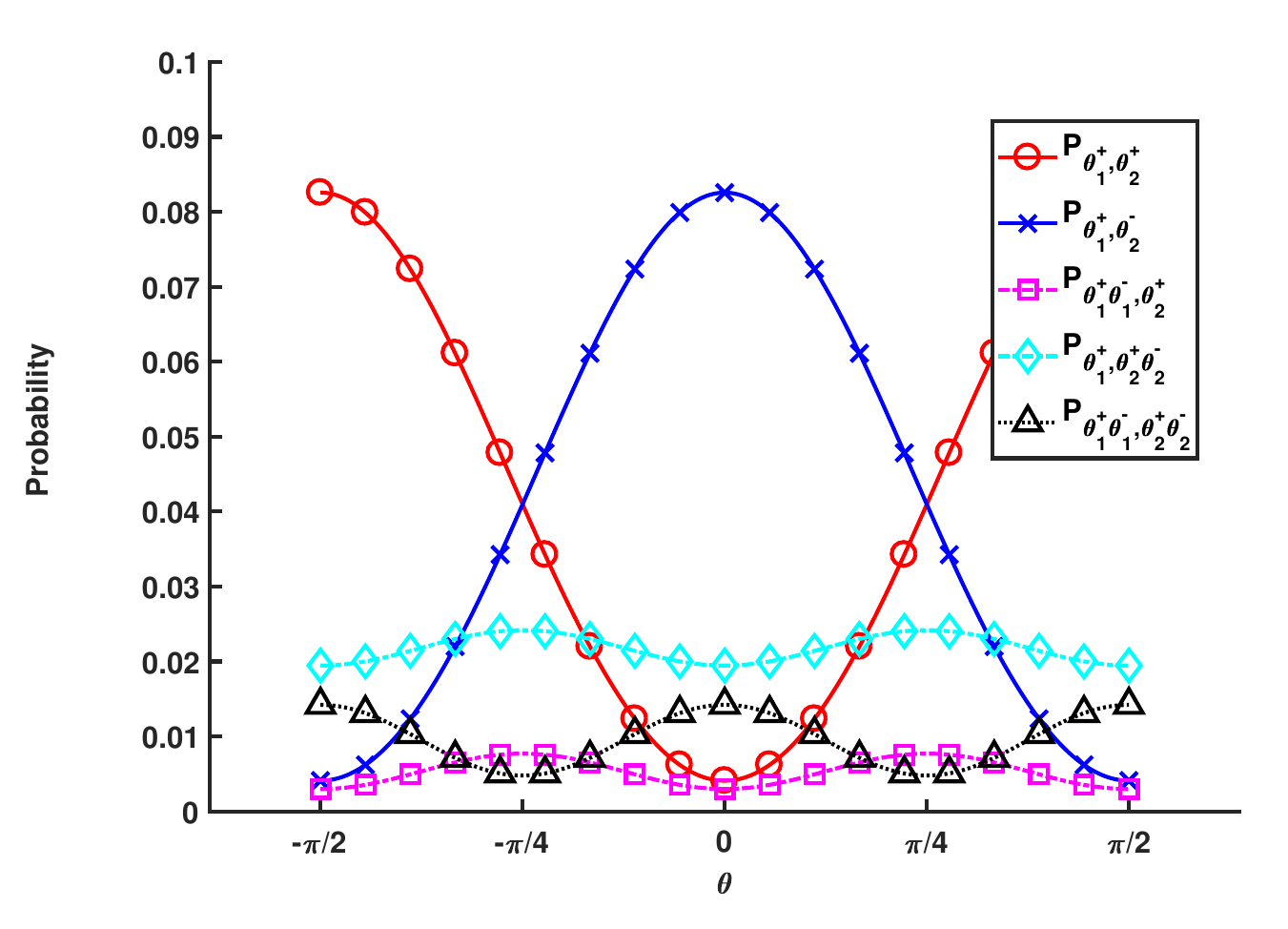}
    \caption{The probabilities for the 16 measurement results given in Table. (\ref{tab:prob}) are plotted as functions of $\theta$. This plot corresponds to the case where $\tau_1=0.7$, $\tau_2=0.3$, and $g=0.6$. The measurement values can be calculated analytically using Eq. \ref{eq:Q-func}.}
    \label{fig:theta}
\end{figure}

While nonlocality of quantum states is not a necessary condition for the security of the BBM92 protocol, the results of a Bell test can provide some insights about the quantum information system. We observe how the CHSH value changes with the intensity of the HBEPP. First, Alice and Bob prepare their measurement bases at angles of $\varphi_{A1}=0^\circ$, $\varphi_{A2}=45^\circ$, $\varphi_{B1}=22.5^\circ$, and $\varphi_{B2}=67.5^\circ$, respectively. The CHSH value is given as 
\begin{subequations}
\begin{eqnarray}\label{eq:chsh}
    S &=& |E(\varphi_{A1},\varphi_{B1})-E(\varphi_{A1},\varphi_{B2}) \nonumber\\ &&+E(\varphi_{A2},\varphi_{B1})+E(\varphi_{A2},\varphi_{B2})|,
\end{eqnarray}
\begin{equation}\label{eq:estimate}
    E(\varphi_A,\varphi_B)
    =\frac{N_{++}(\theta)-N_{+-}(\theta)-N_{-+}(\theta)+N_{--}(\theta)}
    {N_{++}(\theta)+N_{+-}(\theta)+N_{-+}(\theta)+N_{--}(\theta)}.
\end{equation}
\end{subequations}

When adopting the Squash model\cite{beaudry2008Squashing}, The coincidence rates between Alice and Bob are calculated as $N_{\pm\pm}(\theta) \equiv P_{\theta_1^\pm,\theta_2^\pm} + \frac{1}{2}P_{\theta_1^+\theta_1^-,\theta_2^\pm} + \frac{1}{2}P_{\theta_1^\pm,\theta_2^+\theta_2^-} + \frac{1}{4}P_{\theta_1^+\theta_1^-,\theta_2^+\theta_2^-}$. Assuming that all data is stored by time stamping via FPGA, there is an alternative method to discard N-fold coincidence events by post-selecting only two-fold coincidences as valid coincidences. We will refer to this for convenience as the \textit{Discard model}. However, the Discard model cannot overlook the possibility of a loophole regarding data acquisition based on the experimental performer's intention. Therefore, in double-click cases where both detectors of Alice (or Bob) register clicks, the CHSH value should be analyzed using the Squash model method. In this method, the FPGA randomly assigns a value of 0 or 1 with equal probability.
\begin{figure}[t]
    \centering
    \includegraphics[width=\linewidth]{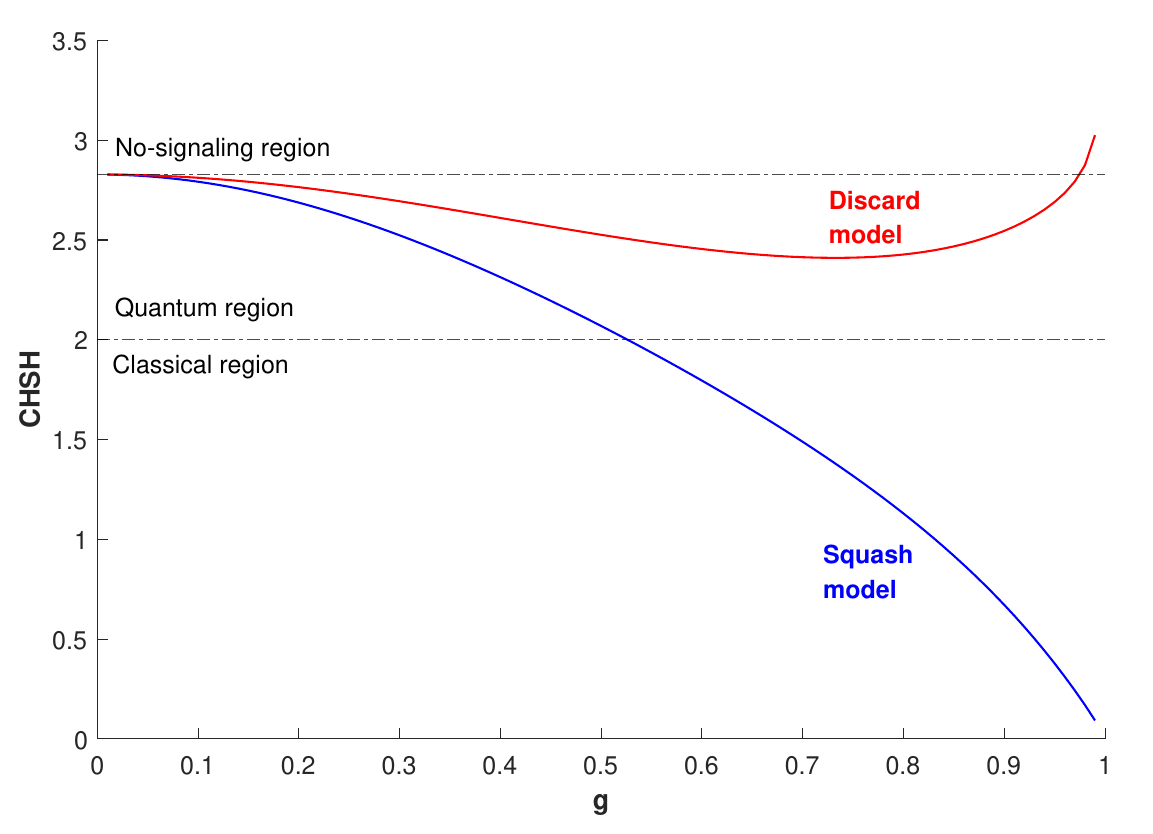}
    \caption{The CHSH value is analyzed for the Squash model and the Discard model with respect to the nonlinear gain $g$. The transmittance values for Alice and Bob are set to $\tau_1=0.7$ and $\tau_2=0.01$, respectively. The red solid line represents the Discard model, while the blue solid line represents the Squash model.}
    \label{fig:chsh}
\end{figure}

The reason why the Squash model must be used in conducting Bell tests is clearly shown in Fig. \ref{fig:chsh}. In the Squash model, as the nonlinear gain $g$ increases, the multi-photon effect of the entangled photon pairs becomes more significant. This results in a higher probability of error, which causes the CHSH value to continuously decrease. This is a perfectly natural and intuitive result. In contrast, for the Discard model, the CHSH value initially decreases gradually but begins to increase again after a certain point. Remarkably, this even exceeds the boundary value of $S=2\sqrt{2}$ at which the quantum state can be maximally exhibited. It appears that we could experimentally reach the No-signaling region, but of course, this is not the case. Although the Discard model claims to eliminate data that are obviously erroneous, it ultimately results in a fake violation because the experimenter's intention intervenes in the data rejection process. This behavior is said to occur because the reconstructed quantum state after data rejection is not positive semidefinite \cite{semenov2011fake}. The fake violations in the Bell test do not directly impact the security of QKD, but they suggest that the widely used entanglement distillation protocol (EDP)-based security proofs may not be applicable. In fact, while the security of BBM92 based on the Squash model has been proven \cite{tsurumaru2008security, kravtsov2023security}, the security of BBM92 based on the Discard model requires additional careful review \cite{koashi2008security}. This is expected to have interesting implications for the advancement of BBM92.

\begin{table*}[t]
    \centering
    \caption{Experimental parameters for SQKD using HBEPP sources \cite{yin2017satellite, yin2017satelliteSci}. Given the center wavelength of $810\text{nm}$, Alice's loss is set to be $1.6\text{dB}$. We note that the coincidence window is shorter than the coherent time $t_c$, emphasizing that the given values of $d_x$ and $\bar{\mu}$ satisfy the single temporal mode condition.}
    \renewcommand{\arraystretch}{1.5} 
    \begin{tabular*}{\textwidth}{@{\extracolsep\fill}cccccc@{\extracolsep\fill}}
    \hline\hline
    \begin{tabular}[c]{@{}c@{}}\textbf{Wavelength}\\ \textbf{{[}nm{]}}\end{tabular} & \begin{tabular}[c]{@{}c@{}}\textbf{Coincidence window}\\ \textbf{{[}ns{]}}\end{tabular} & \begin{tabular}[c]{@{}c@{}}\textbf{Coherent time: $t_c$}\\ \textbf{{[}ns{]}}\end{tabular} & \begin{tabular}[c]{@{}c@{}}\textbf{Loss of Alice: $L_1$}\\ \textbf{{[}dB{]}}\end{tabular} & \begin{tabular}[c]{@{}c@{}}\textbf{Darkcount rate: $d_x$}\\ \textbf{{[}counts per $t_c${]}}\end{tabular} & \begin{tabular}[c]{@{}c@{}}\textbf{Brightness: $\bar{\mu}$}\\ \textbf{{[}pairs per $t_c${]}}\end{tabular} \\ \hline
    $810$ & $2.0$ & $6.25$  & $1.6$ & $6.25\times 10^{-7}$ & $0.037$                                                                    \\ \hline\hline
    \end{tabular*}\label{tab:yin2017}
\end{table*}

\begin{figure}
    \centering
    \includegraphics[width=\linewidth]{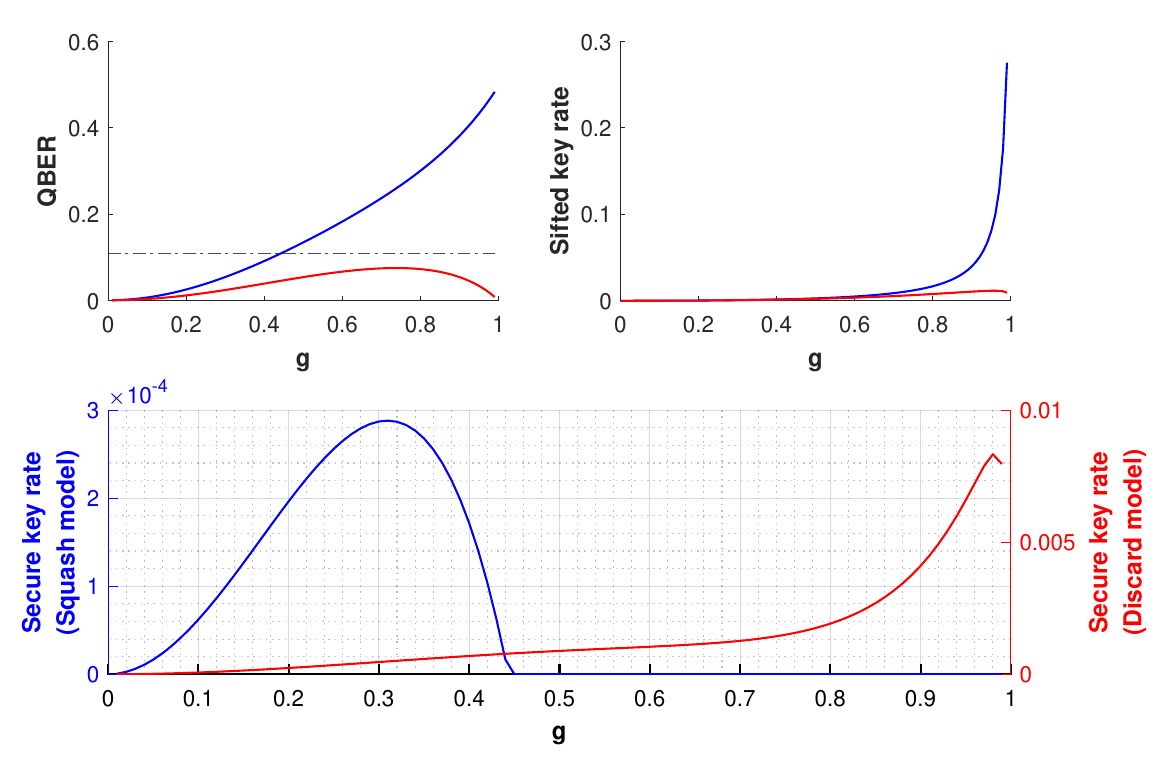}
    \caption{The graph shows QBER, sifted key rate, and secure key rate as functions of nonlinear gain $g$, specifically for the case where $\tau_1=0.7$ and $\tau_2=0.01$. In particular, this paper analyzes the scenario when using a continuous wave pump; therefore, the sifted key rate and secure key rate represent the probability values of sifted key and secure key that can occur during the interaction time $T_{int}$. The red solid line represents the case using the Discard model, while the blue solid line represents the case using the Squash model.}
    \label{fig:Rsec}
\end{figure}

The choice of the Squash model and Discard model as methods for handling double-click events does not only affect the value of CHSH. Naturally, it also influences QBER and key rate in the BBM92 system. Assuming that error correction and privacy amplification are performed as post-processing steps in BBM92 \cite{gottesman2004security}, the secure key rate is given by 
\begin{equation}
    R_{sec}\leq R_{sift}\left[
    1-(1+f(\epsilon))H_2(\epsilon)
    \right],\label{eq:Rsec}
\end{equation}
where $R_{sift}$ is the sifted key rate, $\epsilon$ is the QBER, $f(\epsilon)$ is the error correction efficiency, and $H_2(\epsilon)$ is the Shannon entropy function. The error correction efficiency is a function of the error rate satisfying $f(\epsilon) \leq 1$, we analyzed using the Shannon limit, $f(\epsilon) = 1$. The Shannon entropy function is given by $H_2(\epsilon) = -\epsilon \log_2(\epsilon) - (1-\epsilon) \log_2(1-\epsilon)$. Originally, in Eq. (\ref{eq:Rsec}), bit-flip errors and phase errors should be separated; fortunately, under the conditions of this paper, the bit-flip error and phase error are the same at $\epsilon$.

From Fig. \ref{fig:Rsec} using the Squash model, both the QBER and the sifted key rate increase as the nonlinear gain $g$ rises. This happens because a larger mean photon number increases the key rate between Alice and Bob. However, this also introduces a trade-off, as the error rate rises due to the multi-photon effect. The Discard model exhibits a pattern similar to that of CHSH, where QBER initially increases and then decreases. On the other hand, discarding double-click events results in a relatively low sifted key rate. Nevertheless, when comparing secure key rates, it is consistently found that the secure key rate of the Discard model is higher than that of the Squash model. This aspect indicates that the Discard model is valuable for identifying the range of nonlinear gain $g$ values where security can be achieved, performing a more efficient QKD. However, we note that our research will only focus on the Squash model. 
\begin{figure}
    \centering
    \includegraphics[width=\linewidth]{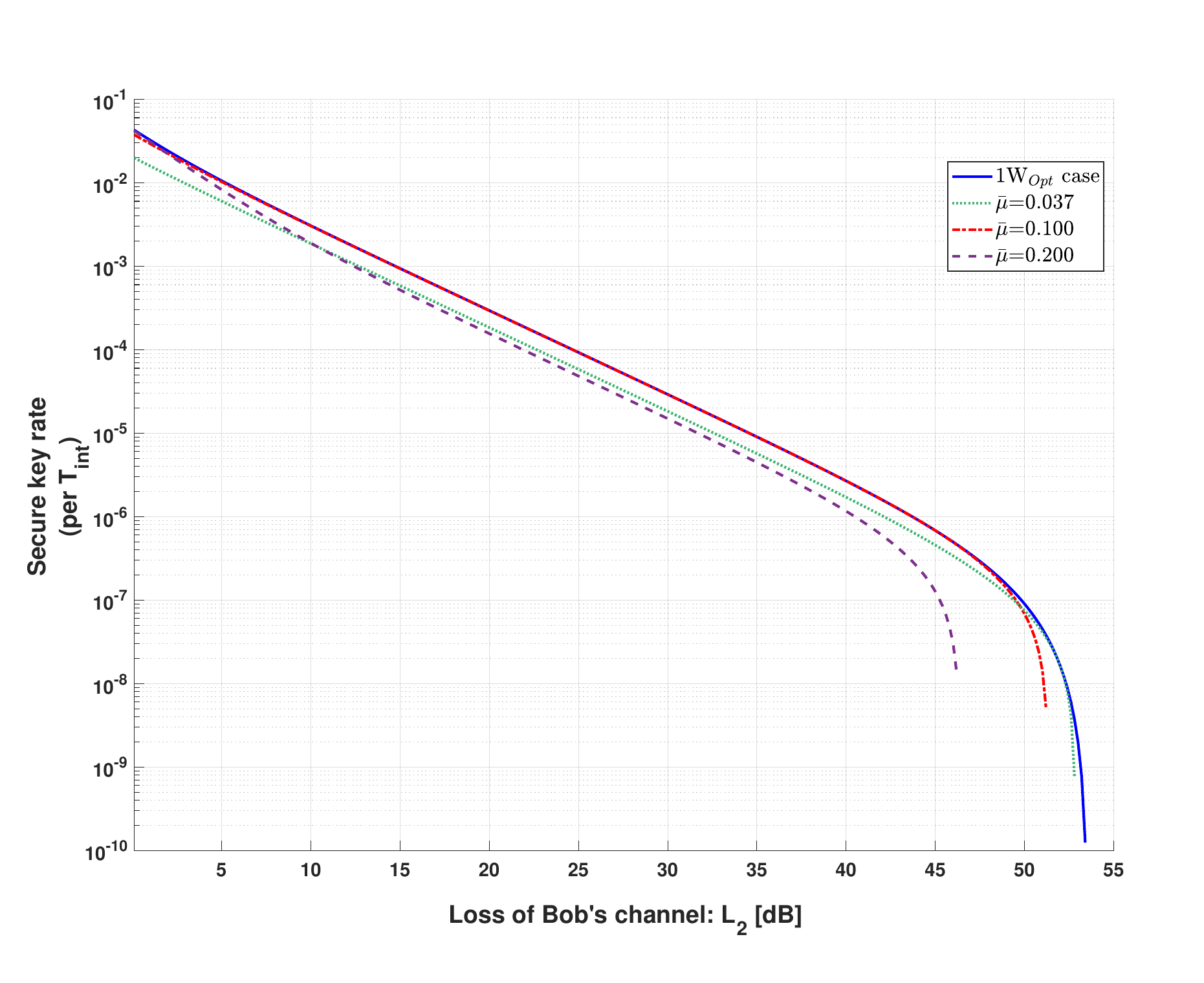}
    \caption{The secure key rate in a one-way communication entanglement-based SQKD system as a function of Bob's loss. The key rate refers to the bit rate during the interaction time that satisfies the single temporal mode condition, rather than the bit rate over one second. The blue solid line represents the key rate when optimizing the intensity of the source according to Bob's channel loss $L_2=-10\log_{10}{\tau_2}$; the red dashed line represents the key rate when we propose $\bar{\mu}=0.1$; the purple dashed line represents the key rate when $\bar{\mu}=0.2$. The green dashed line is based on experimental values given in Table (\ref{tab:yin2017}).}
    \label{fig:keyloss}
\end{figure}
\newline\indent In the SQKD system, in the case of a one-way communication system where Alice has an HBEPP source in the satellite and sends photons to Bob at the ground station, Alice experiences almost no loss. At the same time, Bob incurs a significant loss of about 20 to 45 dB. In a two-way communication system where the entangled photon pair source is located on the satellite and both Alice and Bob are at separate ground stations, both would experience significant losses. To analyze cases of asymmetrical loss, we calculate the optimized nonlinear gain $g_{opt}$ such that the secure key rate is maximized. In this case, the mean photon number at $g_{opt}$ is given by $\bar{\mu}=\sinh^2{(\tanh^{-1}{(g_{opt})})}$. When considering the one-way communication system, it was confirmed that when Alice's loss is sufficiently small, the value of the average number of photons does not significantly change in the region where Bob's loss is between 20 to 45 dB at $\bar{\mu}=0.1$.

To achieve the maximum key rate while conducting the QKD protocol, it is necessary to optimize the mean photon number of the source according to the corresponding channel loss. This consideration also applies to SQKD, where the distance between the satellite and the ground station changes in real-time. However, varying the intensity of the HBEPP source in the SQKD system in real-time can be somewhat burdensome. There are two methods for adjusting the intensity of the source. The first is by varying the current to change the pump beam's intensity. While convenient, this approach alters the laser's wavelength and affects the entangled photon pairs' properties \cite{steinlechner2014efficient}. The second method is using an attenuator to adjust the source's intensity. However, this method requires additional components, which can still be burdensome due to the need for a compact and low-power operation of the satellite module.

We find that the optimized value of the mean photon number required over the range of channel loss in a one-way communication SQKD system remains largely consistent. Therefore, we propose a passive-intensity BBM92 using a HBEPP source with a mean photon number of $\bar{\mu}=0.1$. To assess the effectiveness of the proposed SQKD scheme, we analyzed changes in the secure key rate using results from actual SQKD studies \cite{yin2017satellite, yin2017satelliteSci}. The conditions of the experiment are summarized in Table (\ref{tab:yin2017}). Due to the lack of information on the transmittance value for Alice's receiver located on the satellite, and considering that the performance of SPAD in the wavelength range of $810\text{nm}$ can reach approximately $70\%$, we set Alice's loss to $1.6\text{dB}$. Using the presented parameter values, the secure key rate with respect to Bob's channel loss is given in Fig. (\ref{fig:keyloss}).

Our analysis of the secure key rate shows that the passive-intensity BBM92 in the channel loss range where SQKD primarily operates ensures $99.7\%$ performance compared to the ideal case. This emphasizes that there is no need to optimize the intensity of the HBEPP source at the expense of increasing risks and costs in the SQKD system. Of course, this does not mean that the intensity of the source can be set arbitrarily. In fact, the source used in the experiment had a mean photon number of $\bar{\mu}=0.037$ \cite{yin2017satelliteSci}, which results in a performance of only $62.5\%$ compared to the ideal case when $L_2=20\text{dB}$, and $66\%$ when $L_2=45\text{dB}$. Additionally, in the case of $\bar{\mu}=0.2$, it shows that excessive multi-photon effects from the entangled photon pairs degrade the QKD performance.

\section{Conclusions}
In conclusion, we demonstrate that the measurement probabilities for an HBEPP source distributed through an asymmetric loss channel can be calculated analytically. Based on this, we propose utilizing a passive-intensity HBEPP source in a one-way communication SQKD system. We validated our results by demonstrating outcomes similar to those obtained using the Fusimi-Kano method \cite{brewster2021quantum}.Furthermore, in quantum information processing using HBEPP sources and threshold detectors, double-click events should be addressed using the Squash model rather than the Discard model, which has been shown to produce false violations \cite{semenov2011fake}. It implies the need for the Squash model in BBM92 protocols \cite{tsurumaru2008security, kravtsov2023security}. Our analysis of entangled pair distribution indicated that the optimized intensity of the source remains relatively constant within the operational range of one-way communication SQKD. We found that even with a fixed mean photon number of $\bar{\mu}=1$, the performance was $99.7\%$ of the ideal BBM92 protocol, supporting the use of passive-intensity sources for HBEPP. The proposed idea is expected to yield at least a $60\%$ performance improvement compared to the BBM92 experimental setup conducted in 2017 \cite{yin2017satellite, yin2017satelliteSci}. Although our analysis focuses on one-way communication SQKD, we believe the proposed method is also applicable to two-way communication systems \cite{cao2018bell} and can be valuable for any application involving HBEPP source distribution systems.

\appendix
\section{Probabilities for 16 cases}
In this section, we describe how to calculate the 16 probabilities that can occur when high-brightness entangled photon pairs are distributed to Alice and Bob. For ease of calculation, we have redefined the function $Q_j(\theta)$, which follows the form of Eq. (\ref{eq:Q-func}). When $A,B\in\left\{1,1-\tau_1\right\}$ and $C,D\in\left\{1,1-\tau_2\right\}$ as indicated in Table. (\ref{tab:app_P}), the 16 probability values take the form of a linear combination of $Q_j(\theta)$ as given in Eq. (\ref{eq:app_P}).

\begin{subequations}
    \begin{equation}
        P_{0,0} = (1-g^2)^2\cdot Q_1(\theta),
    \end{equation}
    \begin{equation}
        P_{\theta_1^+,0} = (1-g^2)^2\cdot Q_2(\theta) - P_{0,0},
    \end{equation}
    \begin{equation}
        P_{\theta_1^-,0} = (1-g^2)^2\cdot Q_3(\theta) - P_{0,0},
    \end{equation}
    \begin{equation}
        P_{0,\theta_2^+} = (1-g^2)^2\cdot Q_4(\theta) - P_{0,0},
    \end{equation}
    \begin{equation}
        P_{0,\theta_2^-} = (1-g^2)^2\cdot Q_5(\theta) - P_{0,0},
    \end{equation}
    \begin{equation}
        P_{\theta_1^+,\theta_2^+} = (1-g^2)^2\cdot Q_6(\theta) 
        - \left(P_{\theta_1^+,0} + P_{0,\theta_2^+}\right)
        - P_{0,0},
    \end{equation}
    \begin{equation}
        P_{\theta_1^+,\theta_2^-} = (1-g^2)^2\cdot Q_7(\theta) 
        - \left(P_{\theta_1^+,0} + P_{0,\theta_2^-}\right)
        - P_{0,0},
    \end{equation}
    \begin{equation}
        P_{\theta_1^-,\theta_2^+} = (1-g^2)^2\cdot Q_8(\theta) 
        - \left(P_{\theta_1^-,0} + P_{0,\theta_2^+}\right)
        - P_{0,0},
    \end{equation}
    \begin{equation}
        P_{\theta_1^-,\theta_2^-} = (1-g^2)^2\cdot Q_9(\theta) 
        - \left(P_{\theta_1^-,0} + P_{0,\theta_2^-}\right)
        - P_{0,0},
    \end{equation}
    \begin{equation}
        P_{\theta_1^+\theta_1^-,0} = (1-g^2)^2\cdot Q_{10}(\theta) 
        - \left(P_{\theta_1^+,0} + P_{\theta_1^-,0}\right)
        - P_{0,0},
    \end{equation}
    \begin{equation}
        P_{0,\theta_2^+\theta_2^-} = (1-g^2)^2\cdot Q_{11}(\theta) 
        - \left(P_{0,\theta_2^+} + P_{0,\theta_2^-}\right)
        - P_{0,0},
    \end{equation}
    \begin{eqnarray}
        P_{\theta_1^+\theta_1^-,\theta_2^+} 
        &=& (1-g^2)^2\cdot Q_{12}(\theta) 
        - \left(P_{\theta_1^+,\theta_2^+}+P_{\theta_1^-,\theta_2^+}+
        P_{\theta_1^+\theta_1^-,0}\right)\nonumber\\
        &&- \left(P_{\theta_1^+,0}+P_{\theta_1^-,0}+P_{0,\theta_2^+}
        \right)
        - P_{0,0},
    \end{eqnarray}
    \begin{eqnarray}
        P_{\theta_1^+\theta_1^-,\theta_2^-} 
        &=& (1-g^2)^2\cdot Q_{13}(\theta) 
        - \left(P_{\theta_1^+,\theta_2^-}+P_{\theta_1^-,\theta_2^-}+
        P_{\theta_1^+\theta_1^-,0}\right)\nonumber\\
        &&- \left(P_{\theta_1^+,0}+P_{\theta_1^-,0}+P_{0,\theta_2^-}
        \right)
        - P_{0,0},
    \end{eqnarray}
    \begin{eqnarray}
        P_{\theta_1^+,\theta_2^+\theta_2^-} 
        &=& (1-g^2)^2\cdot Q_{14}(\theta) 
        - \left(P_{\theta_1^+,\theta_2^+}+P_{\theta_1^+,\theta_2^-}+
        P_{0,\theta_2^+\theta_2^-}\right)\nonumber\\
        &&- \left(P_{\theta_1^+,0}+P_{0,\theta_2^+}+P_{0,\theta_2^-}
        \right)
        - P_{0,0},
    \end{eqnarray}
    \begin{eqnarray}
        P_{\theta_1^-,\theta_2^+\theta_2^-} 
        &=& (1-g^2)^2\cdot Q_{15}(\theta) 
        - \left(P_{\theta_1^-,\theta_2^+}+P_{\theta_1^-,\theta_2^-}+
        P_{0,\theta_2^+\theta_2^-}\right)\nonumber\\
        &&- \left(P_{\theta_1^-,0}+P_{0,\theta_2^+}+P_{0,\theta_2^-}
        \right)
        - P_{0,0},
    \end{eqnarray}
    \begin{eqnarray}
        P_{\theta_1^+\theta_1^-,\theta_2^+\theta_2^-} 
        &=& (1-g^2)^2\cdot Q_{16}(\theta) 
        - \left(P_{\theta_1^+\theta_1^-,\theta_2^+}
        +P_{\theta_1^+\theta_1^-,\theta_2^-}
        +\cdots\right.\nonumber\\
        &&\left. P_{\theta_1^+,\theta_2^+\theta_2^-}
        +P_{\theta_1^-,\theta_2^+\theta_2^-}\right) 
        - \left(P_{\theta_1^+,\theta_2^+}
        +P_{\theta_1^+,\theta_2^-}+\cdots\right.\nonumber\\
        && \left. P_{\theta_1^-,\theta_2^+}+
         P_{\theta_1^-,\theta_2^-} +P_{\theta_1^+\theta_1^-,0}+P_{0,\theta_2^+\theta_2^-}\right)\nonumber\\
        && - \left(P_{\theta_1^+,0}+P_{\theta_1^-,0}
        +P_{0,\theta_2^+}+P_{0,\theta_2^-}\right)
        - P_{0,0}.
    \end{eqnarray}\label{eq:app_P}
\end{subequations}

\begin{table}[]
\caption{Each $Q_j(\theta)$ functions are given with Eq. (\ref{eq:Q-func}) when there coefficients $A,B,C,D$ are given as follows.} 
\renewcommand{\arraystretch}{1.3} 
\resizebox{\columnwidth}{!}{
\begin{tabular}{c cc cc c c cc cc}
\hline\hline
\multirow{2}{*}{\textbf{$Q_j(\theta)$}} & \multicolumn{2}{c}{\textbf{Alice}}          & \multicolumn{2}{c}{\textbf{Bob}}            &  & \multirow{2}{*}
{\textbf{$Q_j(\theta)$}} & \multicolumn{2}{c}{\textbf{Alice}}          & \multicolumn{2}{c}{\textbf{Bob}}            \\ \cline{2-5} \cline{8-11} 
                                      & \multicolumn{1}{c}{\textbf{A}} & \textbf{B} & \multicolumn{1}{c}{\textbf{C}} & \textbf{D} &  &                                       & \multicolumn{1}{c}{\textbf{A}} & \textbf{B} & \multicolumn{1}{c}{\textbf{C}} & \textbf{D} \\ \cline{1-5} \cline{7-11} 
\textbf{$Q_1(\theta)$}                           
& \multicolumn{1}{c}{$1$} & $1$  & \multicolumn{1}{c}{$1$} & $1$          &  & \textbf{$Q_9(\theta)$}                           
& \multicolumn{1}{c}{1} & $1-\tau_1$ & \multicolumn{1}{c}{$1$} & $1-\tau_2$ \\[1.2ex] \cline{1-5} \cline{7-11} 
\textbf{$Q_2(\theta)$}                           & \multicolumn{1}{c}{$1-\tau_1$}          & $1$         & \multicolumn{1}{c}{$1$}          & $1$          &  & \textbf{$Q_{10}(\theta)$}                           & \multicolumn{1}{c}{$1-\tau_1$}          & $1-\tau_1$          & \multicolumn{1}{c}{$1$}          & $1$          \\[1.2ex] \cline{1-5} \cline{7-11} 
\textbf{$Q_3(\theta)$}                           & \multicolumn{1}{c}{$1$}          & $1-\tau_1$          & \multicolumn{1}{c}{$1$}          & $1$          &  & \textbf{$Q_{11}(\theta)$}                          & \multicolumn{1}{c}{$1$}          & $1$          & \multicolumn{1}{c}{$1-\tau_2$}          & $1-\tau_2$          \\[1.2ex] \cline{1-5} \cline{7-11} 
\textbf{$Q_4(\theta)$}                           & \multicolumn{1}{c}{$1$}          & $1$          & \multicolumn{1}{c}{$1-\tau_2$}          & $1$          &  & \textbf{$Q_{12}(\theta)$}                          & \multicolumn{1}{c}{$1-\tau_1$}          & $1-\tau_1$          & \multicolumn{1}{c}{$1-\tau_2$}          & $1$          \\[1.2ex] \cline{1-5} \cline{7-11} 
\textbf{$Q_5(\theta)$}                           & \multicolumn{1}{c}{$1$}          & $1$          & \multicolumn{1}{c}{$1$}          & $1-\tau_2$          &  & \textbf{$Q_{13}(\theta)$}                          & \multicolumn{1}{c}{$1-\tau_1$}          & $1-\tau_1$          & \multicolumn{1}{c}{$1$}          & $1-\tau_2$          \\[1.2ex] \cline{1-5} \cline{7-11} 
\textbf{$Q_6(\theta)$}                           & \multicolumn{1}{c}{$1-\tau_1$}          & $1$          & \multicolumn{1}{c}{$1-\tau_2$}          & $1$         &  & \textbf{$Q_{14}(\theta)$}                          & \multicolumn{1}{c}{$1-\tau_1$}          & $1$          & \multicolumn{1}{c}{$1-\tau_2$}          & $1-\tau_2$          \\[1.2ex] \cline{1-5} \cline{7-11} 
\textbf{$Q_7(\theta)$}                           & \multicolumn{1}{c}{$1-\tau_1$}          & $1$          & \multicolumn{1}{c}{$1$}          & $1-\tau_2$          &  & \textbf{$Q_{15}(\theta)$}                          & \multicolumn{1}{c}{$1$}          & $1-\tau_1$          & \multicolumn{1}{c}{$1-\tau_2$}          & $1-\tau_2$          \\[1.2ex] \cline{1-5} \cline{7-11} 
\textbf{$Q_8(\theta)$}                           & \multicolumn{1}{c}{$1$}          & $1-\tau_1$          & \multicolumn{1}{c}{$1-\tau_2$}          & $1$          &  & \textbf{$Q_{16}(\theta)$}                          & \multicolumn{1}{c}{$1-\tau_1$}          & $1-\tau_1$          & \multicolumn{1}{c}{$1-\tau_2$}          & $1-\tau_2$         \\[1.2ex]
\hline\hline
\end{tabular}
}\label{tab:app_P}
\end{table}
\section*{Acknowledgments}
This research was supported by the Challengeable Future Defense Technology Research and Development Program through the Agency For Defense Development(ADD) funded by the Defense Acquisition Program Administration(DAPA) in 2023(No.915027201). 

\bibliography{ldmpref}

\end{document}